# Synthesis of $CaKFe_4As_4$ bulk samples with high critical current density using a spark plasma sintering technique

**Shigeyuki Ishida[1], S. Pavan Kumar Naik[1], Yoshinori Tsuchiya[1], Yasunori Mawatari[1], Yoshiyuki Yoshida[1], Akira Iyo[1], Hiroshi Eisaki[1], Yoshihisa Kamiya[1,2], Kenji Kawashima[1,2] and Hiraku Ogino[1]**

[1]Research Institute for Advanced Electronics and Photonics, National Institute of Advanced Industrial Science and Technology (AIST), Umezono 1-1-1, Tsukuba, Ibaraki 305-8568, Japan
[2]IMRA Materials R&D Co., Ltd., Asahi-machi 2-1, Kariya, Aichi 448-0032, Japan

**Abstract**

A high density $CaKFe_4As_4$ bulk sample was successfully synthesized using a spark plasma sintering (SPS) technique. The density of the synthesized sample was 5.02 g cm$^{-3}$, corresponding to 96.2% of the theoretical density of $CaKFe_4As_4$. Moreover, a reasonably high Vickers hardness of 1 GPa was measured. The electrical resistivity of the SPS bulk sample was as low as approximately 600 μΩ cm at 300 K, which is smaller than that of the ordinary sintered polycrystalline sample by nearly one order of magnitude, and exhibited a sharp superconducting transition, with the transition width $\Delta T_c$ less than 2 K, indicating an improved grain connectivity. The critical current density of the SPS bulk sample, as calculated from the magnetization hysteresis loops (magnetic $J_c$), reached 18 kA cm$^{-2}$ at 4.2 K under 5 T, which is the highest among the iron-based superconductor polycrystalline samples reported thus far.

---

## 1. Introduction

Iron-based superconductors (IBSs) possess relatively high superconducting transition temperature ($T_c$) exceeding 50 K, as well as large and nearly isotropic upper critical field ($H_{c2}$), i.e. about 100 T at 4.2 K. Due to these characteristics, IBSs are considered promising candidates for the high magnetic field applications, either in the form of wire-wound magnets or as trapped field magnets [1-6].

Thus far, various efforts have been made to fabricate IBS wires, tapes, and bulk samples. Generally, it is recognized that the critical current density ($J_c$) of the polycrystalline samples is determined by their grain boundary (GB) characteristics [7,8]. According to previous studies, the inter-grain $J_c$ is limited by (i) impurity phases, (ii) voids and cracks, and (iii) grain-to-grain misorientations. Regarding the factors (i) and (ii), non-superconducting impurity phases and/or voids/cracks block super-currents resulting in the reduction of inter-grain $J_c$. Especially, the short coherence length and the low carrier density of IBSs tend to make the problem more serious. Factor (iii) is related to the anisotropic superconducting properties due to the layered structure as well as to the unconventional (non-$s_{++}$) pairing symmetry [7]. Additionally, it has been demonstrated that the misalignment of grains along the *c*-axis and within the *ab*-plane deteriorates inter-grain $J_c$. Therefore, in order to achieve high $J_c$, one needs to synthesize impurity-free, high-density, and highly textured samples.

To obtain high-density bulk samples, Weiss *et al.* employed a hot isostatic press (HIP) technique. They encapsulated pre-reacted $Ba_{1-x}K_xFe_2As_2$ powders, wrapped with Ag foil into a steel tube, and heated them at 600 °C under the pressure of 193 MPa [9,10]. The density of the resultant bulk samples was as large as, approximately, 98% of the ideal value, and the Vickers hardness was about 3.5 GPa, suggesting that the samples were nearly free from voids/cracks. Indeed, the recorded $J_c$ in their samples was 110 kA cm$^{-2}$ at 4.2 K under self-field, which is the highest among the reported values. For reference purposes, the magnetic $J_c$ values of representative IBS bulk samples are listed in table 1.

Apart from HIP, the spark plasma sintering (SPS) technique is known as a useful technique to obtain dense bulk samples through a simple procedure [16]. Different from HIP and other heating systems that use external heat sources, SPS uses pulsed currents that pass through conductive dies (typically made of graphite) as well as samples themselves that produce Joule heating. The internal heat generation, allowing us to shorten sintering time and lower the sintering temperature, is effective in preventing grain coarsening. Thus far, the SPS technique has been used for fabricating $MgB_2$ bulks, yielding dense samples with around 99% of theoretical density, resulting in high $J_c$ values [17,18]. As for IBSs, SPS has been applied to the synthesis of $NdFeAsO_{0.75}F_{0.25}$ [19] and $FeSe_{0.5}Te_{0.5}$ [20], and dense bulk samples with a high relative density of 96% and 90%, respectively, were synthesized successfully. As a result, it is worth trying to apply the SPS method to other IBSs.

Here, we report the synthesis of a $CaKFe_4As_4$ bulk sample using the SPS technique. $CaKFe_4As_4$ has the same crystal structure as $Ba_{1-x}K_xFe_2As_2$, whereas the Ca : K ratio is fixed to be 1:1 [21]. Its superconducting properties ($T_c$ and $H_{c2}$) are comparable to those of optimally doped $Ba_{1-x}K_xFe_2As_2$ [22]. It should be noted that $CaKFe_4As_4$ single crystalline samples exhibit excellent critical current properties especially at relatively high temperatures, i.e. $J_c$ at 20 K under 3 T is approximately 1 MA cm$^{-1}$ and 10 MA cm$^{-1}$ for *H* // *c*-axis and *ab*-plane, respectively, which are 10 (for *H* // *c*-axis) to 100 (for *H* // *ab*-plane) times larger than those for $Ba_{1-x}K_xFe_2As_2$ [23-26]. In this study, we demonstrate that the bulk sample fabricated using SPS method has a high density, reaching 96% of theoretical density. Moreover, it has a high $J_c$ of 18 kA cm$^{-2}$ at 4.2 K under 5 T, which is the highest $J_c$ value among those of IBS bulk samples reported so far.

## 2. Methods

First, the polycrystalline $CaKFe_4As_4$ sample was synthesized using a solid-state reaction. Throughout the process, materials were handled in a nitrogen-filled glove box. Prior to the synthesis of the sample, CaAs, KAs, and $Fe_2As$ precursors were synthesized. The starting reagents - Ca (3 N, chunks), K (3 N, chunks), Fe (4 N, 150 mesh powders), and As (6 N, 1–5 mm grains) - were weighed in the appropriate molar ratio. The typical weights were 10 g (CaAs), 20 g (KAs), and 30 g ($Fe_2As$), respectively. Then, the mixture was sealed in a quartz tube (CaAs and $Fe_2As$) or a stainless-steel container (KAs) and heated for 10 h at 750 °C (CaAs), 650 °C (KAs), and 850 °C ($Fe_2As$). Afterwards, the reacted precursors were weighed and mixed with the ratio of CaAs : KAs : $Fe_2As$, which was equal to 1:1.05:1. Here, taking into account the loss of K and As during the reaction process, 0.05 more of KAs was added. The mixture (around 4 g in total) was pelletized and wrapped in a tantalum (Ta) foil and sealed in a stainless-steel container. The container was heated at 900 °C for 3 h and rapidly cooled in cold water down to room temperature (RT). The rapid cooling was effective to avoid the formation of $CaFe_2As_2$ and $KFe_2As_2$ which occurs at temperature range of 800-900 °C owing to the decomposition of $CaKFe_4As_4$. The sample was re-milled, pelletized under approximately 10 MPa, wrapped in a Ta foil and sealed in a stainless-steel container again. Then, the container was heated at 920 °C for 3 h, followed by rapid cooling to RT. Hereafter, the bulk sample that was synthesized at ambient pressure is called "AP bulk". A part of the sample was cut into a rectangular shape and used for various physical property measurements, and the other part was ground into powder and used for the following SPS process.

First, the $CaKFe_4As_4$ powder (approximately 1 g) was inserted into a graphite die with an inner diameter of 10 mm and placed in a SPS apparatus (SS Alloy Co., Ltd.). A uniaxial pressure of 50 MPa was applied to the graphite die. The sample chamber was evacuated and then, filled with argon gas to a pressure of 0.5 atm. The graphite die was heated to 700-800 °C with a rate of 50 °C $min^{-1}$ by applying current pulses. The advantage of SPS technique is that the densification finishes in a short time (typically 5-10 min), which effectively reduces the formation of $CaFe_2As_2$ and $KFe_2As_2$. The densification process was followed by a heat treatment for 2 h at 500 °C where $CaKFe_4As_4$ was stable, and then, the sample was furnace cooled to RT. The sintered bulk (named "SPS bulk") was cut into a rectangular shape for physical property measurements.

X-ray diffraction (XRD) of the powder was performed using CuKα radiation (Rigaku, Ultima IV), equipped with a high-speed detector system (Rigaku, D/teX Ultra). Magnetization measurements were performed using a SQUID magnetometer (Quantum Design, MPMS). Electrical resistivity measurements were carried out through a four-probe method using silver pastes (Dupont, 4922N) for contacts. Vickers hardness ($H_V$) was measured on the polished surface of the bulk sample using a Vickers microhardness testing machine (Matsuzawa MMT-X3). Ten different positions were measured with a load of 200 g and a duration of 10 s, and the average value was calculated. The microstructure was observed and compositional analysis was performed using a scanning electron microscope (SEM) (Carl Zeiss, ULTRA55) equipped with an energy dispersive x-ray spectrometer (EDS) (Thermo Fisher Scientific, NSS312E). For the measurements, the bulk surface was polished through Ar ion milling using a cross-section polisher (Gatan, PECS II Model 685).

## 3. Results

### *3.1. Look of the sample and XRD patterns*

A photograph of the SPS bulk (10 mm in diameter and 2.3 mm in thickness, with a weight of 0.9 g) is shown in the inset of figure 1(b). A shiny surface was obtained after polishing the bulk using abrasive papers and lapping films. The density of the SPS bulk was 5.02(1) g $cm^{-3}$, which corresponds to 96.2(2)% of the theoretical density of $CaKFe_4As_4$ (5.22 g $cm^{-3}$). This value is much higher than that of the AP bulk (3.87(1) g $cm^{-3}$ that is equal to 74.1(1)% of theoretical density) and comparable with that of a dense $Ba_{0.6}K_{0.4}Fe_2As_2$ bulk fabricated using HIP [10]. Thus, we successfully obtained a $CaKFe_4As_4$ bulk with high density by a simple procedure using the SPS technique. For the mechanical property, the average $H_V$ of the SPS bulk was 0.99(2) GPa. This value is lower than the $H_V$ (3.5 GPa) of the $Ba_{0.6}K_{0.4}Fe_2As_2$ HIP bulk [9], whereas it is reasonably high as a bulk sample without outer hard metals (such as steel) considering that the $H_v$ values for IBS wire/tape cores fabricated using a pure Ag sheath are typically 1-1.5 GPa [27,28]. Note that the $H_v$ values of wires/tapes are known to be enhanced by using harder sheath materials such as stainless steel [29,30]. Then, the higher $H_v$ of the $Ba_{0.6}K_{0.4}Fe_2As_2$ HIP bulk sample can be associated with the use of the outer steel tube which compresses the bulk sample.

Figures 1(a) and 1(b) show the XRD patterns of the $CaKFe_4As_4$ powder (after the second calcination) and the SPS bulk samples, respectively. As seen in figure 1(a), most of the peaks for the powder sample can be indexed by the $CaKFe_4As_4$ phase, confirming that $CaKFe_4As_4$ is the primary phase. Additionally, small peaks of the $KFe_2As_2$ and $CaFe_2As_2$ phases are identified, as indicated by the squares and triangles, respectively. For the SPS bulk sample, figure 1(b) shows that the main phase is still $CaKFe_4As_4$, and the peak intensities of $KFe_2As_2$ and $CaFe_2As_2$ phases are not enhanced, demonstrating that the formation of those phases can be avoided by applying SPS technique. On the other hand, several additional peaks are observed as indicated by stars, which are identified as FeAs. Since these peaks from the FeAs phase are negligibly small for the original powder sample, the decomposition of the $CaKFe_4As_4$ phase is likely to have occurred during the SPS process. Moreover, there is no appreciable difference in the peak intensities of $CaKFe_4As_4$ between the XRD patterns of powder and SPS bulk samples. This indicates that the grain alignment did not occur during the SPS process.

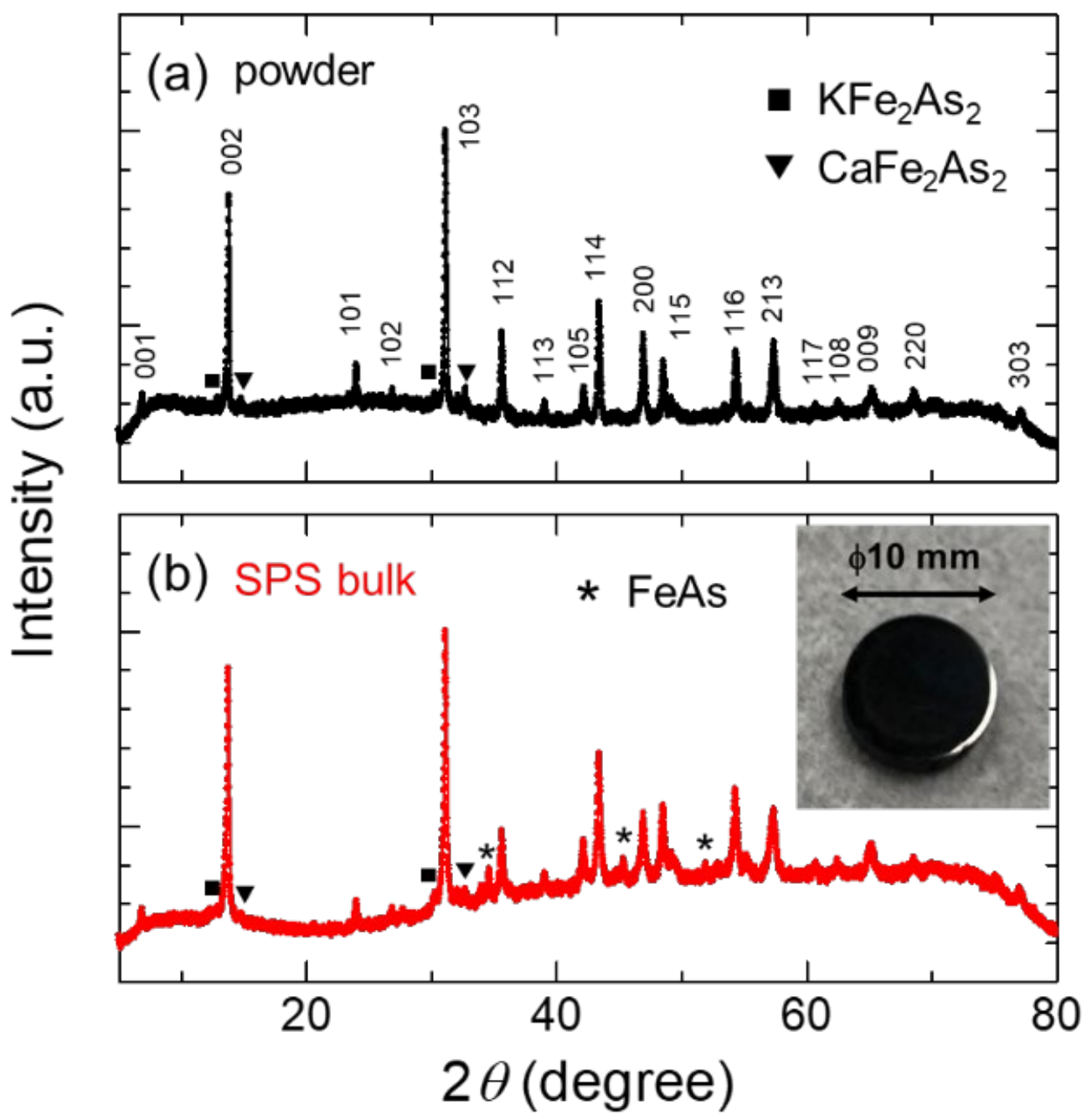

**Figure 1.** XRD patterns of $CaKFe_4As_4$ powder (a) and SPS bulk (b). Squares, triangles, and stars indicate peaks of $KFe_2As_2$, $CaFe_2As_2$, and FeAs phases, respectively. The inset of (b) shows the photograph of SPS bulk.

*3.2 SEM/EDS analysis*

Figure 2(a) shows the SEM image of the SPS bulk surface. The different brightness is an indication of differences in chemical composition and/or crystal orientation; hence, it is considered to correspond to different grains. Most of the grains show ellipsoidal (or plate-like) shapes with various dimensions in the range of 1-10 μm. The long axes of the grains are randomly oriented, i.e. there is no trace of texturing in the SPS bulk as indicated by the XRD patterns. Moreover, there are some black spots between grains that are indicative of impurity phases or voids. From the surface asperities, in most cases, these black spots are considered to be impurities. Similarly, the black lines at the GBs are considered to be microcracks.

Figure 2(b) shows the results of the EDS mapping of the constituting elements, i.e. Ca, K, Fe, As, and O, taken at the area indicated by a blue square in figure 2(a), which contains regions with different brightness. It can be seen that most of the area contains Ca, K, Fe, As but not O, confirming that the main phase is $CaKFe_4As_4$. By contrast, in the brightest region numbered 1, Ca and K are almost absent, whereas Fe and As exhibit stronger intensities. This indicates that this region corresponds to FeAs as it is also observed in the XRD pattern. The less bright region, numbered 2, also does not contain Ca and K, whereas more Fe and less As are observed, suggestive of $Fe_2As$, despite it not being identified in the XRD pattern. For the region numbered 3, where the brightness is similar to the $CaKFe_4As_4$ phase, less Ca and more K are detected without noticeable difference in Fe and As intensities. This indicates that the grain corresponds to $KFe_2As_2$, which is identified in the XRD pattern. In addition, the black spot numbered 4 contains only Ca and O, indicating the precipitation of CaO impurities. Finally, there is a substantial amount of unknown O-rich impurities in GBs. The results revealed the non-negligible O-contamination in our samples.

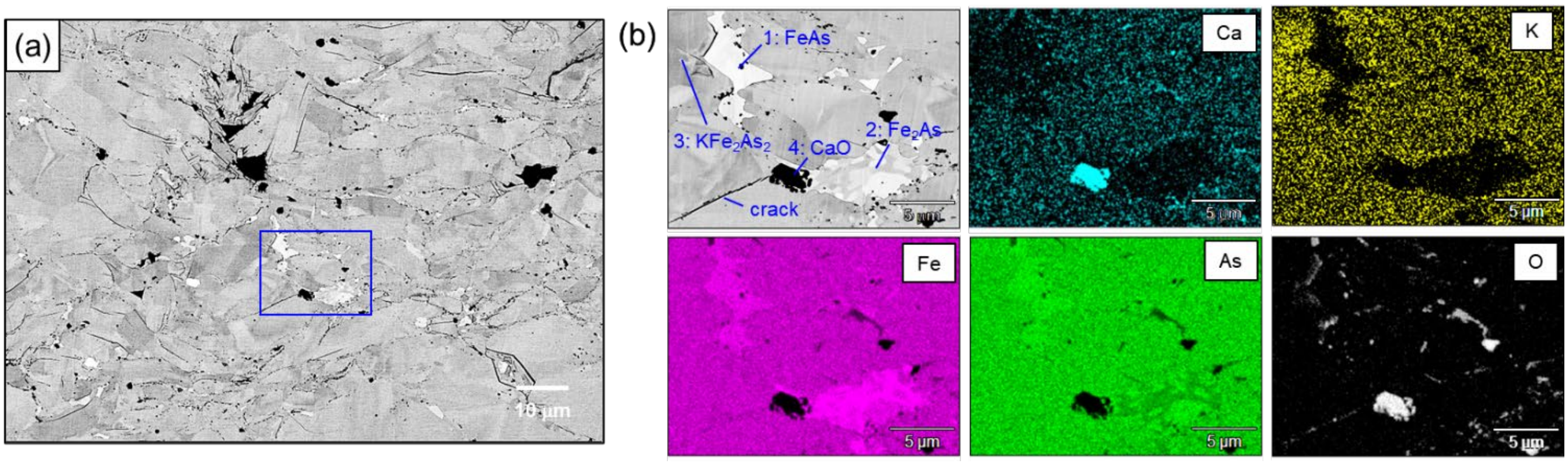

**Figure 2.** (a) SEM image of the cross section of the SPS bulk and (b) mapping of the elements, Ca, K, Fe, As, and O in the area indicated by the blue square in (a).

### *3.3 Superconducting transition temperatures*

The results for the temperature dependence of the magnetic susceptibility ($\chi(T)$) for the SPS (red) and the AP (black) bulk samples are shown in figure 3. Filled and open circles indicate $\chi(T)$ measured in zero-field-cooling (ZFC) and field-cooling (FC) processes, respectively, with an applied field of 10 Oe. Each sample shows a sharp superconducting transition with a large shielding signal, while the SPS bulk shows a sharper transition with $\Delta T_c$ around 2 K, a larger ZFC signal, and a smaller FC signal. The sharper transition and the larger ZFC signal suggest a better electromagnetic homogeneity throughout the SPS bulk sample. Moreover, the smaller FC signal is suggestive of better pinning properties in the SPS bulk. In addition, the $T_c$ values, defined as the initial drop from the normal state value (i.e. $M \sim 0$), are 35.6 K for the SPS bulk and 34.3 K for the AP bulk. The $T_c$ value of the SPS bulk is close to that of the single crystalline $CaKFe_4As_4$ ($T_c$ = 35.7 K) [26,31].

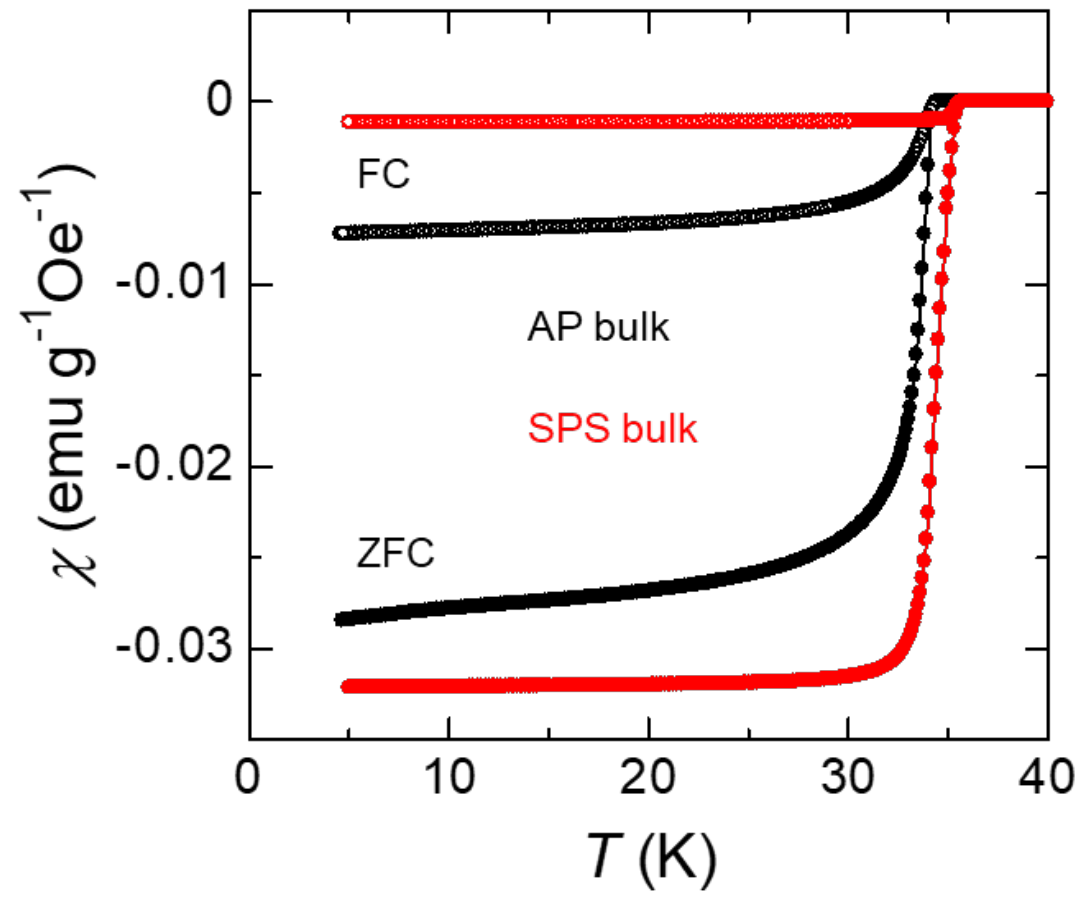

**Figure 3.** Temperature dependence of magnetic susceptibility measured in ZFC (filled circles) and FC (open circles) processes for the SPS bulk (red) and AP bulk (black) with applied magnetic field of 10 Oe.

*3.4 Transport properties*

Figure 4(a) shows the results for the temperature dependence of the electrical resistivity ($\rho(T)$) of the SPS bulk (red) and the AP bulk (black) samples. For the purpose of comparison, $\rho_{ab}(T)$ for $CaKFe_4As_4$ single crystal [26] (blue), which represents a sample with ideal grain connection and alignment, is also plotted. It should be noted that, since the grains in the bulk samples are randomly oriented, the $\rho(T)$ is also affected by $\rho(T)$ along the $c$ axis ($\rho_c(T)$), which is larger than $\rho_{ab}(T)$ ($\rho_c$(300 K) = 1000-2000 μΩ cm [22]). The magnitude of $\rho$(300 K) is 614 μΩ cm for the SPS bulk and 5310 μΩ cm for the AP bulk. In both cases, the $\rho(T)$ shows a metallic behaviour and decreases to 66.5 μΩ cm and 987 μΩ cm at 40 K ($\rho$(40 K)) for the SPS and AP bulks, respectively. The magnitude of $\rho$ of the SPS bulk is smaller by an order of magnitude than that of the AP bulk, and much closer to that of the single crystalline sample ($\rho_{ab}$(300 K) = 349 μΩ cm and $\rho_{ab}$(40 K) = 25.8 μΩ cm). The smaller $\rho$ observed for the SPS bulk indicates the improved grain connectivity compared with the AP bulk.

In addition, the $\rho(T)$ behaviour below the $T_c$ can also be associated with the grain connectivity. Figure 4(b) shows the magnified view around $T_c$. For the AP bulk, a finite $\rho$ remains at $T_c^{zero}$ = 22.3 K, which is much lower than the $T_c$ determined from $\chi(T)$, thereby indicating the weak-link nature of GBs. By contrast, the $\rho(T)$ of the SPS bulk exhibits a sharp drop around $T$ = 35.3 K, and the zero resistance is observed at $T_c^{zero}$ = 34.9 K. This difference also derives from a better grain connectivity of the SPS bulk.

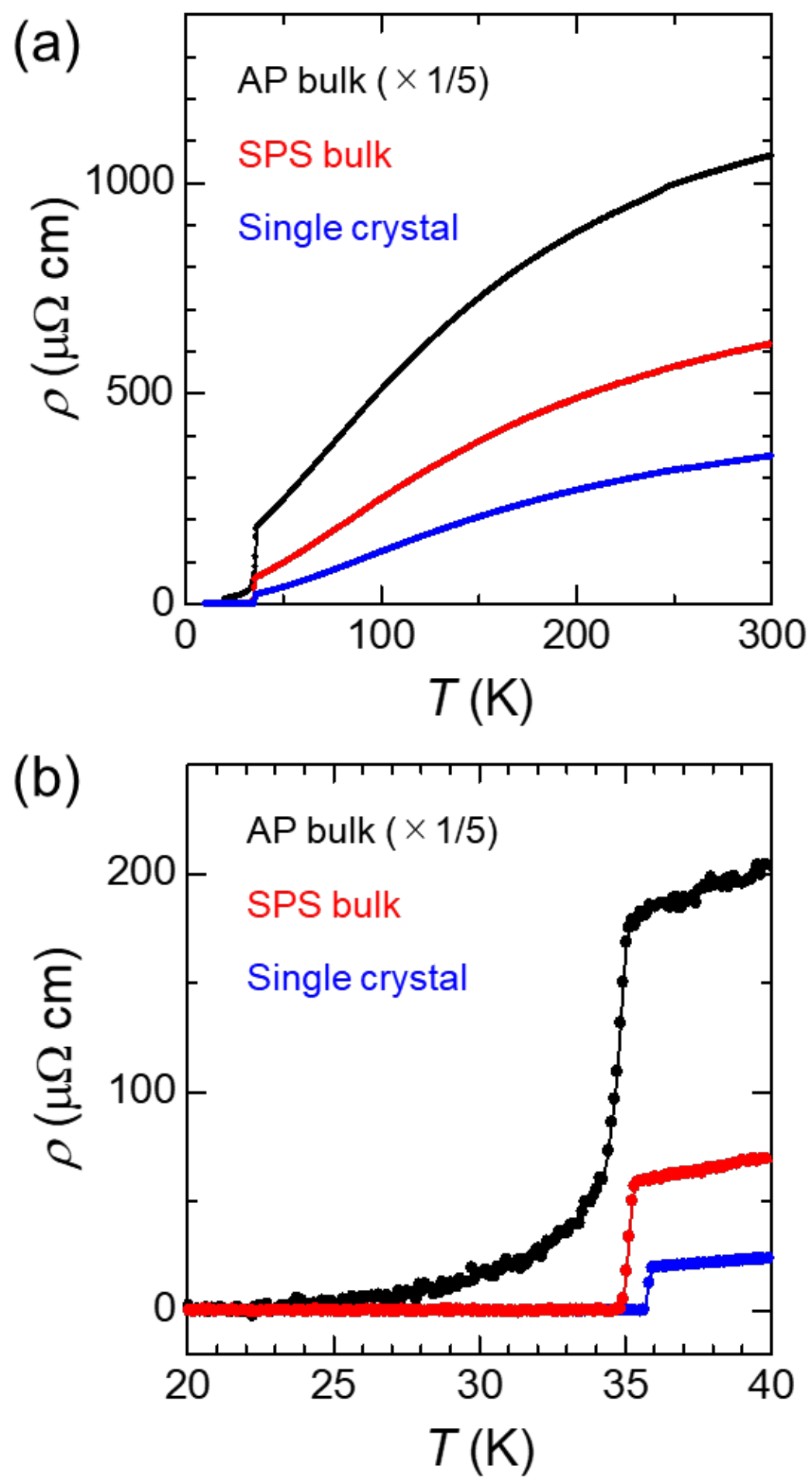

**Figure 4.** (a) Temperature dependence of electrical resistivity for the SPS bulk (red), AP bulk (black) and single crystal [26] (blue). For the AP bulk, $\rho$ is divided by 5. (b) Magnified view around the $T_c$.

*3.5 Critical current properties*

Figures 5(a) and 5(b) show the magnetization hysteresis loops (MHLs) measured at several temperatures (ranging from 4.2 to 30 K) for the SPS bulk and the AP bulk, respectively. In this case, the length ($l$), width ($w$), and thickness ($t$) of the samples were 2.1 mm, 1.2 mm and 1.1 mm for the SPS bulk, whereas for the AP bulk, they were 3.0 mm, 1.2 mm and 0.9 mm, respectively. The magnetic field was applied to the widest plane of the bulk. Each MHL of the SPS bulk shows an almost symmetric shape against the $M$ and $H$ axes. However, the MHL of the AP bulk exhibits an asymmetric shape, which is often observed when MHL is dominated by the intra-grain $J_c$ contribution, i.e. the grain connectivity is weak [32]. The shape of the MHL confirms the improved grain connectivity in the SPS bulk.

When using the MHLs, the magnetic $J_c$ was calculated based on the Bean's critical state model [33], i.e. $J_c = 20\Delta M/w(l - w/3l)$, where $\Delta M$ is the width of the MHLs. Figures 5(c) and 5(d) show the magnetic field dependence of $J_c(H)$ for the SPS bulk and the AP bulk derived from figures 5(a) and 5(b), respectively. The $J_c(H)$ shows a qualitative difference between the two samples in that the one of the SPS bulk decreases with increasing magnetic field whereas that of the AP bulk is characterized by the second peak effect, which is clearly seen for $J_c(H)$ at 30 K (grey data in figure 5(b)). The second peak effect is often observed for various

IBS single crystals [34,35], including $CaKFe_4As_4$ [23-26]. This also suggests that the $J_c(H)$ of the AP bulk is governed by the intra-grain $J_c$ contribution. The $J_c$ values of the SPS bulk are 81 kA $cm^{-2}$ (self-field) and 18 kA $cm^{-2}$ (5 T) at 4.2 K, i.e. 4 to 5 times larger than those of the AP bulk (22 kA $cm^{-2}$ and 2.9 kA $cm^{-2}$ respectively).

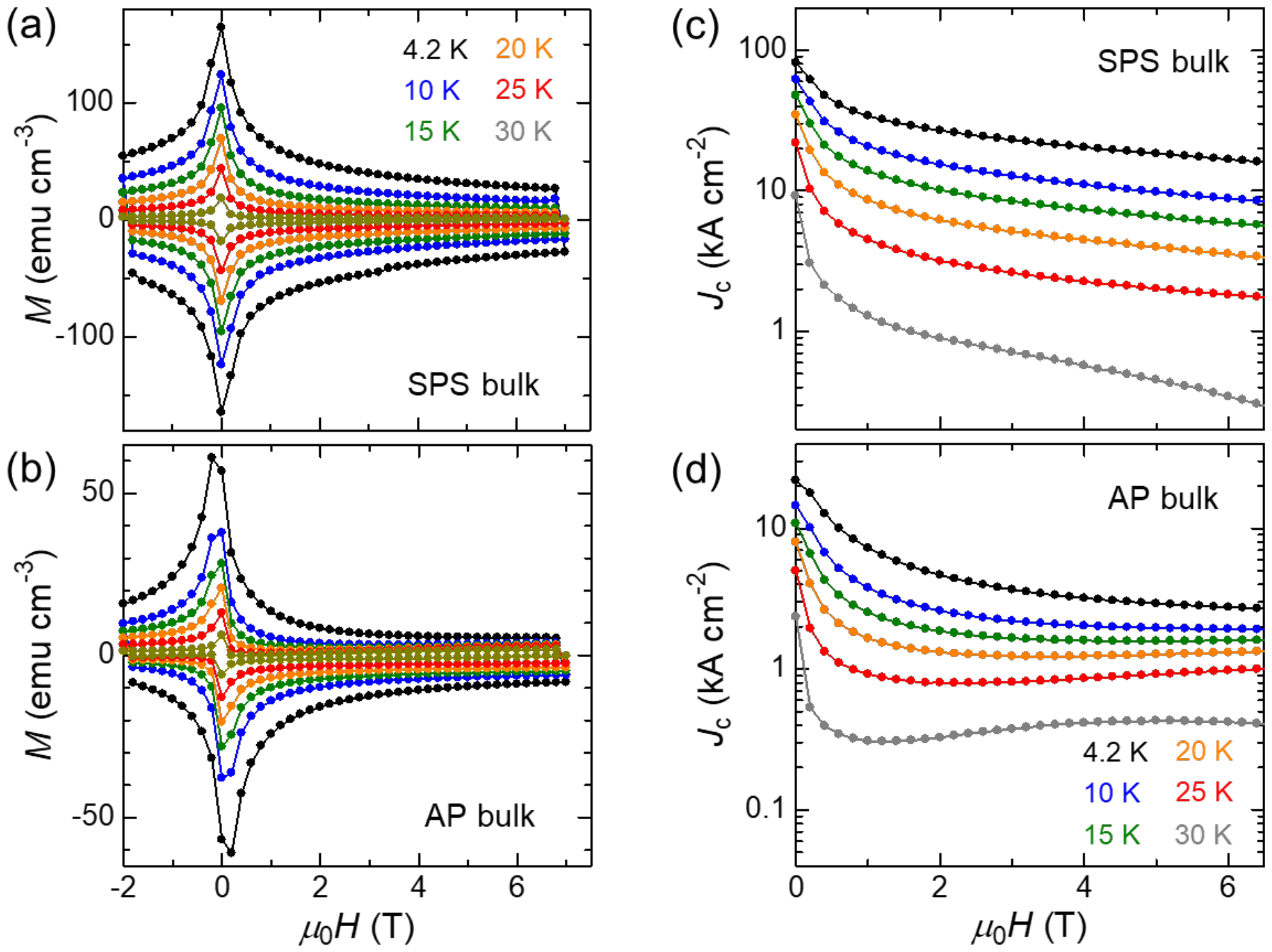

**Figure 5.** Magnetization hysteresis loops (MHLs) for the SPS bulk (a) and AP bulk (b). Magnetic field dependence of $J_c$ derived from MHLs for SPS bulk (c) and AP bulk (d).

## 4. Discussion

In order to compare the $J_c$ values of the present SPS bulk with other IBS polycrystalline bulk samples existing in literature, the magnetic $J_c$ values (at 4.2 K (or 5 K) and 20 K under self-field and 5 T) for $Ba_{1-x}K_xFe_2As_2$ [10,11], $Ba(Fe_{1-x}Co_x)_2As_2$ [11], $CaKFe_4As_4$ [12], $REFeAsO_{1-x}F_x$ (RE = Sm and Nd) [13,14], and $FeSe_{0.5}Te_{0.5}$ [15] (AP and HIP denote the sintering conditions) are listed in table 1 together with the results of this work. Moreover, in figure 6, the temperature dependences of $J_c(T)$ under 5 T are plotted (5 T is chosen because IBSs are considered to be used under high fields). Among the reported bulk samples (open symbols) at 4.2 K (or 5 K), $Ba_{0.6}K_{0.4}Fe_2As_2$ (HIP) exhibits a high $J_c$ value of 12 kA $cm^{-2}$ (red circle). $SmFeAsO_{0.85}$ (AP) (blue reversed triangle), $SmFeAsO_{1-x}F_x$ with In addition (AP) (purple diamond), $Ba_{0.6}K_{0.4}Fe_2As_2$ (HIP) (orange triangle) and $CaKFe_4As_4$ (AP) (green square) show similar $J_c$ values in the range of 7 to 9 kA $cm^{-2}$. $Ba_{0.6}K_{0.4}Fe_2As_2$ (AP) bulk and $FeSe_{0.5}Te_{0.5}$ (AP) bulk exhibit relatively low $J_c$ values of 1.3 kA $cm^{-2}$ (black circle) and 0.6 kA $cm^{-2}$ (black cross), respectively. Notably, the present $CaKFe_4As_4$ SPS bulk (red stars) has the highest $J_c$ (18 kA $cm^{-2}$) among the reported IBS bulks. The results suggest that $CaKFe_4As_4$ is a promising material for bulk magnet applications.

| Family | Material | Magnetic $J_c$ (kA cm$^{-2}$) 4.2 K (* 5 K) 0 T | 4.2 K (* 5 K) 5 T | 20 K 0 T | 20 K 5 T | Reference |
|---|---|---|---|---|---|---|
| 122 | $Ba_{0.6}K_{0.4}Fe_2As_2$ (AP) | 10 | 1.3 | 5.0 | 0.31 | [10] |
| | $Ba_{0.6}K_{0.4}Fe_2As_2$ (HIP) | 110 | 12 | 52 | 3.0 | [10] |
| | $Ba_{0.6}K_{0.4}Fe_2As_2$ (HIP) | 82 | 7.2 | N/A | N/A | [11] |
| | $Ba_{0.4}K_{0.6}Fe_2As_2$ (HIP) | 71 | 5.5 | N/A | N/A | [11] |
| | $Ba(Fe_{0.92}Co_{0.08})_2As_2$ (HIP) | 11 | 0.47 | N/A | N/A | [11] |
| 1144 | $CaKFe_4As_4$ (AP) | 25* | 7.0* | 9.0 | 1.0 | [12] |
| | $CaKFe_4As_4$ (AP) | 22 | 2.9 | 7.9 | 1.3 | [This work] |
| | $CaKFe_4As_4$ (SPS) | 81 | 18 | 34 | 4.1 | [This work] |
| 1111 | $SmFeAsO_{0.85}$ (AP) | 30 | 8.9 | 6.1 | 1.8 | [13] |
| | $NdFeAsO_{0.94}F_{0.06}$ (AP) | 12 | 2.8 | N/A | N/A | [13] |
| | $SmFeAsO_{1-x}F_x$ + In (AP) | 25 | 7.5 | N/A | N/A | [14] |
| 11 | $FeSe_{0.5}Te_{0.5}$ (AP) | 1.5* | 0.60* | N/A | N/A | [15] |

**Table 1.** Magnetic $J_c$ of various iron-based superconducting bulk samples.

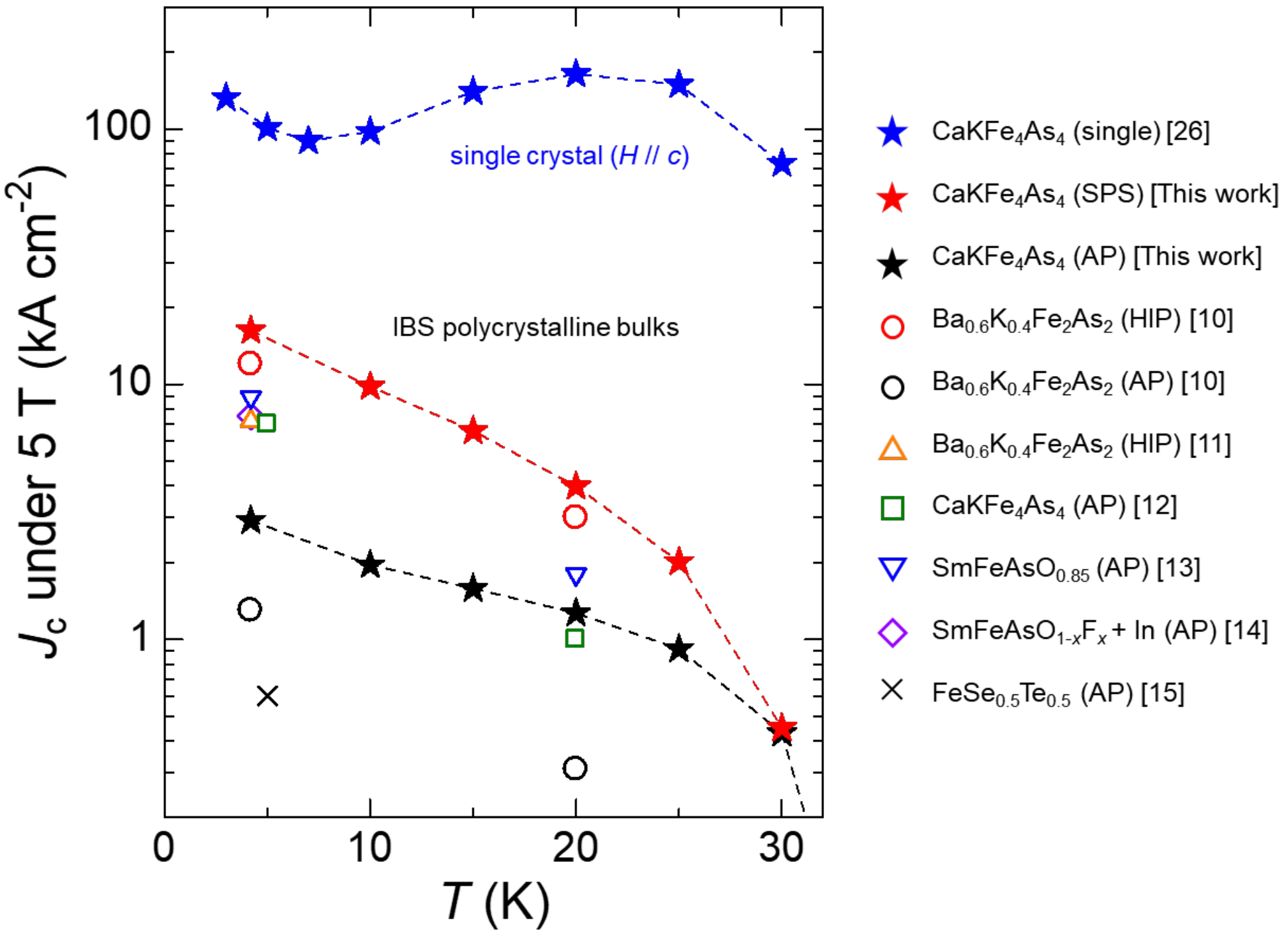

**Figure 6.** Temperature dependence of magnetic $J_c$ under 5 T for $CaKFe_4As_4$ SPS and AP bulks of this study (red and black stars, respectively), various IBS bulks existing in literature [10-15], and $CaKFe_4As_4$ single crystal [26] (blue stars).

In comparison with $CaKFe_4As_4$ single crystal ($H$ // $c$-axis, blue stars) [26], the $J_c$ value of the present SPS bulk is still smaller by one order of magnitude. In addition, $J_c$ shows distinct temperature dependence; $J_c(T)$ of the single crystal is characterized by the broad peak around 20 K, whereas that of the SPS bulk exhibits a monotonous temperature dependence. These results indicate that the $J_c$ of $CaKFe_4As_4$ bulk can be increased by further improving the inter-grain connectivity.

As shown in the XRD pattern (figure 1(b)) and the SEM/EDS analysis (figure 2), the present SPS bulk contains various impurities such as FeAs, $Fe_2As$, $KFe_2As_2$, $CaFe_2As_2$, and oxides at GBs. The grain connectivity will be improved by eliminating these impurity phases. $KFe_2As_2$ and $CaFe_2As_2$ are formed during the synthesis process of the $CaKFe_4As_4$ powder. To minimize the formation of these impurities, it is necessary to optimize the starting chemical composition, synthesis temperature and time, etc. Similarly, to avoid the formation of FeAs, it is essential to tune the SPS synthesis conditions, such as heating rate, sintering temperature, duration time, cooling rate, and applied pressure.

Furthermore, the precipitation of oxide impurities occurs through (i) oxidation of starting reagents, (ii) residual O in the glove box and (iii) residual O in the SPS chamber. The use of high purity starting reagents as well as the reduction of the O level in the glove box is a possible solution to reduce the oxide impurities in the powder/bulk samples. As for the residual O in the SPS chamber, we think that, at this moment, the influence is rather minor because the sample is packed and pressurized in a graphite die.

The SEM image (figure 2(a)) revealed the presence of microcracks, which also cause the reduction of $J_c$. The microcracks possibly originated from the uncompleted densification and/or the thermal shock during the cooling process. It is known that the densification rate of the SPS process strongly depends on the grain size and the packing structure of the initial powder. Thus, grain refining of the initial powder would be meaningful. Moreover, a slow cooling after the densification would be effective if the thermal shock due to the rapid cooling causes cracking.

In this study, we have not obtained any signature for the $c$-axis texture of grains. Considering that the $J_c$ of IBS tapes with textured superconducting cores (e.g. 150 kA $cm^{-2}$ at 4.2 K under 10 T [36]) is much higher than that of the present SPS bulk, texturing should be a key to improve $J_c$ performance of bulk samples. There are already several reports on the successful synthesis of textured bulks for various ceramic samples [16]. For example, a $Ca_3Co_4O_9$ ceramic sample prepared using the SPS technique exhibits a very high density (around 99% of the theoretical density) and a textured structure, resulting in improved thermoelectric properties [37]. Therefore, it is worth trying to modify the SPS process for achieving better grain alignment. Another issue is the relationship between the applied pressure and the microstructure that should be investigated to gain insight into the effects of pressure on texturing. Additionally, the grain refining of the initial powder may have an impact because the grain rotation leading to texturing, generally, depends on the grain size.

## 5. Conclusions

In this study, we fabricated a $CaKFe_4As_4$ bulk by using the SPS technique. It was demonstrated that the SPS

process is efficient for obtaining a dense IBS bulk, where the density was as high as 96% of the theoretical density, and the $H_V$ was 1 GPa, which was reasonably high. The magnetic $J_c$ of the SPS bulk was 18 kA cm$^{-2}$ at 4.2 K under 5 T, which is the highest $J_c$ among those of the reported IBS bulks. Meanwhile, the microstructure analysis revealed a substantial amount of impurity phases and microcracks, which are likely to limit the $J_c$ performance of the present SPS bulk. The results suggest that there is room for improvement of the $J_c$ performance of the $CaKFe_4As_4$ bulk.

## Acknowledgements

This work was supported by the Japan Society for the Promotion of Science (JSPS) Grant-in-Aid for Scientific Research on Innovative Areas and KAKENHI (JSPS Grant Numbers JP16H06439 and 19H02179) and the Hitachi Global Foundation. SPKN would like to thank JSPS for providing PD fellowship (ID: P19354). The Vickers measurement was performed with the technical support of Dr. M. Katoh at AIST. The SEM/EDS measurements were carried out at JFE Techno-Research Corporation with the technical supports by Dr. S. Tsukimoto.

## References

[1] Putti M *et al* 2010 *Superconductor Science and Technology.* **23** 034003

[2] Ma Y W 2012 *Superconductor Science and Technology.* **25** 113001

[3] Shimoyama J 2014 *Superconductor Science and Technology.* **27** 044002

[4] Pallecchi L, Eisterer M, Malagoli A and Putti M 2015 *Superconductor Science and Technology.* **27** 044002

[5] Hosono H, Yamamoto A, Hiramatsu H and Ma Y W 2018 *Materials Today* **21** 278

[6] Yao C and Ma Y 2019 *Superconductor Science and Technology* **32** 023002

[7] Durrell J H et al 2011 *Reports on Progress in Physics* **74** 124511

[8] Iida K, Hänisch J and Yamamoto A 2020 *Superconductor Science and Technology.* **33** 043001

[9] Weiss J D *et al* 2015 *Superconductor Science and Technology* **28** 112001

[10] Weiss J D, Jiang J, Polyanskii A A and Hellstrom E E 2013 *Superconductor Science and Technology* **26** 074003

[11] Kim Y J, Weiss J D, Hellstrom E E, Larbalestier, and Seidman D N 2014 *Applied Physics Letters* **105** 162604

[12] Singh S J *et al* 2020 *Superconductor Science and Technology.* **33** 025003

[13] Yamamoto A *et al* 2008 *Superconductor Science and Technology.* **21** 095008

[14] Fujioka M *et al* 2013 *Journal of Physical Society of Japan.* **82** 024705

[15] Palenzona A *et al* 2012 *Superconductor Science and Technology.* **25** 115018

[16] Orru R, Licheri R, Locci A M, Cincotti A and Cao G C 2008 *Materials Science and Engineering: R:Reports* **63** 127–287

[17] Shim S H, Shim K B and Yoon J W 2005 *Journal of the American Ceramic Society* **88** 858

[18] Noudem J G, Aburras M, Bernstein P, Chaud X, Muralidhar M, and Murakami M 2014 *Journal of*

*Applied Physics* **116** 163916
[19] Kursumovic A, Durrel J H, Chen S K and MacManus-Driscoll J L 2010 *Superconductor Science and Technology* **23** 025022
[20] Puneet P, Podila R, He J, Rao A M, Howard A, Cornell N and Zakhidov A A 2015 *Nanotechnology Reviews* **4** 411-417
[21] Iyo A *et al* 2016 *Journal of the American. Chemical Society* **138** 3410
[22] Meier W R *et al* 2016 *Physical Review B* **94** 064501
[23] Singh S J *et al* 2018 *Physical Review Materials* **2** 074802
[24] Pyon S *et al* 2019 *Physical Review B* **99** 104506
[25] Cheng W., Lin H, Shen B and Wen H H 2019 *Science Bulletin* **64** 31–39
[26] Ishida S *et al* 2019 *npj Quantum Materials* **4** 27
[27] Gao Z, Togano K, Matsumoto A and Kumakura H 2014 *Sci. Rep.* **4** 4065
[28] Lin H *et al* 2014 *Sci. Rep.* **4** 6944
[29] Gao Z, Togano K, Matsumoto A and Kumakura H 2015 *Supercond. Sci. Technol.* **28** 012001
[30] Gao Z, Togano K, Zhang Y, Matsumoto A, Kikuchi A and Kumakura H 2017 *Supercond. Sci. Technol.* **30** 095012
[31] Meier W R, Kong T, Bud'ko S L and Canfield P C 2017 *Physical Review Materials* **1** 013401
[32] Hecher J *et al* 2016 *Superconductor Science and Technology* **29** 025004
[33] Bean C P 1964 *Reviews of Modern Physics* **36** 31
[34] Salem-Sugui S *et al* 2010 *Physical Review B* **82** 054513
[35] Ishida S *et al* 2017 *Physical Review B* **95** 014517
[36] Huang H *et al* 2018 *Supercond. Sci. Technol.* **31** 015017
[37] Liu Y, Lin Y, Shi Z, Nan C W and Shen Z 2005 *Journal of the American Ceramic Society.* **88** 1337-1340